\documentclass[showpacs,aps]{revtex4}
\usepackage{epsfig}

\def\Li2#1{{\mathrm{Li}}_2\left(#1\right)}
\def\ba{\begin{eqnarray}}
\def\ea{\end{eqnarray}}
\def\dd{{\mathrm d}}

\newcommand{\NN}{N}
\newcommand{\DD}{{\mathcal D}}
\newcommand{\NS}{{\mathrm NS}}
\def\order#1{{\mathcal O}\left(#1\right)}

\newlength{\splotwidth}
\begin{document}

\title{
\[ \vspace{-2cm} \]
\noindent\hfill\hbox{\rm Alberta Thy 01-02} \vskip 1pt
\noindent\hfill\hbox{\rm hep-ph/0202102} \vskip 10pt
Muon decay spectrum: leading logarithmic approximation}

\author{Andrej Arbuzov}
\email{aarbuzov@phys.ualberta.ca}
\author{Andrzej Czarnecki}
\email{czar@phys.ualberta.ca}
\author{Andrei Gaponenko}
\email{agapon@phys.ualberta.ca}
\affiliation{Department of Physics, University of Alberta
Edmonton, AB\ \  T6G 2J1, Canada}
\begin{abstract}
$\order{\alpha^2 \ln^2(m_\mu/m_e)}$ QED corrections to the electron
spectrum and angular distribution in muon decay are evaluated.  
Impact on the determination of
Michel parameters is estimated. Current theoretical uncertainty
in the muon decay distributions is discussed.
\end{abstract}
\maketitle

\section{Introduction}

The decay $\mu^+ \to e^+ \nu_e \overline{\nu}_\mu$ has been very
accurately studied.  
Its total rate determines
the Fermi constant $G_\mu$ describing the 
strength of the weak interactions
\cite{Marciano:1999ih}.  Standard Model predicts other features of this
decay, such
as the positron energy and angular distribution, positron polarization,
and a variety of correlations between spins and momenta of the muon
and its decay products
\cite{Sachs72,Scheck:1978yg,Fetscher:1986uj,Fetscher:1993ki}.  Precise
determinations of those observables test the Standard Model and can be
used to search for possible ``new physics'' effects
\cite{Kuno:1999jp}.  This motivates ongoing studies of the
positron transverse \cite{Bodek:2000pv} and
longitudinal polarization \cite{PrieelsQu} and
angular/energy distributions \cite{Rodning:2001js,Quraan:2000vq}.
Muon lifetime is also being remeasured \cite{carey,kirkby}.

Current experiments are so precise that  theoretical predictions must
include radiative corrections beyond the first order in the fine
structure constant
$\alpha\simeq 1/137.036$.
For the muon lifetime they are known
\cite{berman62,Behrends:1956mb,Berman:1958ti,Kinoshita:1959ru,%
vanRitbergen:1998yd,Steinhauser:1999bx,Czarnecki:2001cz} through
$\order{\alpha^2}$.  On the other hand, little is known about
the positron energy distribution beyond the $\order{\alpha}$
effects \cite{Kinoshita:1959ru,Arbuzov:2001ui}.  

The experiment TWIST at Canada's National Laboratory TRIUMF is
designed to measure the positron spectrum from polarized $\mu^+$
decays with a precision of $10^{-4}$
\cite{Rodning:2001js,Quraan:2000vq}.  To match this, and help search for
``new physics'' effects, the Standard Model prediction must include at
least the leading $\order{\alpha^2}$ effects in the positron
distribution.  

We have examined the dominant effects at this order, which arise due
to emission of collinear photons and $e^+e^-$ pairs.  Such effects do
not significantly affect the total muon lifetime, but they rearrange
the positron spectrum.
They are enhanced by two powers of the muon and electron mass ratio
logarithm $L \equiv  \ln (m_\mu^2/m_e^2) \simeq 10.66$ (this
roughly determines  their order of magnitude, $(\alpha^2/
\pi^2) L^2 \simeq 6\cdot 10^{-4}$).

We are interested in the energy and angular distribution of positrons
produced in the $\mu^+$ decay.  We normalize it so that it coincides
with the differential width of the decay in the two
lowest orders in $\alpha$.  A difference occurs at $\order{\alpha^2}$
where an additional positron can appear due to pair production.
The differential distribution of positrons (summed over $e^+$ spin states) 
in the polarized muon decay is
\ba \label{spectrum}
\frac{\dd^2\NN^{\mu^+ \to e^+
        \nu\bar{\nu}}}{\dd z \,\, \dd \cos\theta} 
&=& \Gamma_0\left[ F(z) - \cos\theta \, P_\mu G(z) \right],
\qquad
\Gamma_0 \equiv \frac{G_\mu^2 m_\mu^5}{192\pi^3}
\biggl( 1 + \frac{3}{5}\;\frac{m_\mu^2}{m_W^2} \biggr)\, , 
\nonumber \\
z &\equiv & {2E\over (1+r^2)m_\mu}\, , 
 \qquad r \equiv \frac{m_e}{m_\mu}\, ,
\qquad z_0 \le z \le 1, \qquad 
z_0 \equiv \frac{2r}{1+r^2}\, ,
\ea
where $m_e$ and $m_\mu$ are the electron and muon masses; $\theta$ is
the angle between the positron flight direction and the muon spin (for
the $\mu^-$ decay, the sign of the $\cos\theta$ term should be reversed);
$P_{\mu}$ is the degree of the muon polarization; $E$ is the positron
energy.

Functions $F(z)$ and $G(z)$ describe the isotropic and anisotropic
parts of the positron spectrum.  They can be expanded in series in
$\alpha$, 
\ba
F(z) = f_{\rm Born}(z) + {\alpha\over 2\pi} f_1(z)
 + \left( {\alpha\over 2\pi} \right)^2 f_2(z)
 + \order{\alpha^3},
\ea
and similarly for $G(z)$.  

To match the precision of TWIST, the electron mass should be included
at the Born level,
\ba
f_{\rm Born}(z) &=& 6(1+r^2)^4 \, vz
 \biggl[ z(1-z) 
+ \frac{2}{9}\rho ( 4z^2 - 3z - z_0^2)
+ \eta  z_0(1-z) \biggr],
\nonumber \\
g_{\rm Born}(z) &=& -2 (1+r^2)^4 \, v^2  z^2 \xi \biggl[ 1 - z
+ \frac{2}{3}\delta \left(4z - 3 -r z_0\right) \biggr],
\nonumber \\
v &\equiv & \sqrt{1-\frac{m_e^2}{E^2}}, 
\ea
where $\rho$, $\eta$, $\xi$, and $\delta$ are the so--called Michel
parameters~\cite{mich1,mich2,mich3} which depend on the Lorentz
structure of the interaction responsible for the decay.  In this paper
we assume that the decay is caused by the Standard Model $V-A$
interaction, for which $\rho =3/4$, $\eta =0$, $\xi =1$, and $\delta
=3/4$. These values agree with present experimental fits
\cite{Groom:2000in}, 
\ba \label{pdf}
\rho        &=& 0.7518  \pm 0.0026, 
\nonumber \\
\eta        &=& - 0.007 \pm 0.013,
\nonumber \\
\xi P_{\mu} &=& 1.0027  \pm 0.0079 \pm 0.0030,
\nonumber \\
\delta      &=& 0.7486  \pm 0.0026 \pm 0.0028.
\ea

In the massless limit $(m_e \to 0)$ we have
\ba
f_{\rm Born}(z) 
&\to & f_0(z) = z^2(3 - 2z), \qquad
g_{\rm Born}(z) 
\to 
g_0(z) = z^2(1 - 2z).
\label{eqzero}
\ea

Functions $f_1$ \cite{Kinoshita:1959ru} and $g_1$
\cite{Arbuzov:2001ui} are also known with full dependence on the electron
mass.  However, at present it is sufficient to use their massless
limit given in the Appendix.

The $\order{\alpha^2}$ effects are not yet known.  They can be divided
up into three parts according to the power of $L$,
\ba
f_2(z) ={(L-1)^2\over 2} f_2^{\rm LL}(z)  + (L-1)  f_2^{\rm NLL}(z)
 + \Delta f_2(z),
\ea
and similarly for $g_2$.  In this paper we evaluate the leading
logarithmic (LL) corrections $f_2^{\rm LL}$ and
$g_2^{\rm LL}$. We divide them into contributions
of pure photon emissions and of diagrams with $e^+e^-$ pairs,
\ba
f_2^{\rm LL} \equiv f_2^{{\rm LL}(\gamma)} + f_2^{{\rm LL}(e^+e^-)},\qquad
g_2^{\rm LL} \equiv g_2^{{\rm LL}(\gamma)} + g_2^{{\rm LL}(e^+e^-)}.
\label{fullLL}
\ea
In diagrams with $e^+e^-$ pairs we have to clarify the meaning of the
variable $z$, whether it describes the energy of the ``primary''
positron or the one from the pair.  (We can neglect interference in the LL
approximation.)  These two possibilities give rise to the so--called
non-singlet (NS) and singlet (S) parts of pair corrections,
\ba
f_2^{{\rm LL}(e^+e^-)} \equiv 
{2\over 3} f^{{\rm LL}(e^+e^-)}_{2\, {\rm NS}}
+f^{{\rm LL}(e^+e^-)}_{2\, {\rm S}},
\qquad
g_2^{{\rm LL}(e^+e^-)} \equiv 
{2\over 3} g^{{\rm LL}(e^+e^-)}_{2\, {\rm NS}}
+g^{{\rm LL}(e^+e^-)}_{2\, {\rm S}}.
\ea
Ingredients needed to evaluate the full LL effect in eq.~(\ref{fullLL})
are given below in eqs.~(\ref{f2gamma}, \ref{g2gamma}) 
(photonic corrections) and (\ref{ns2ee}, \ref{g2ee}) (pairs).

\section{Leading Logarithmic Approximation}

The LL
corrections can be found by convoluting 
the tree level spectrum (\ref{eqzero})
with the positron structure function (SF), a solution of the
Dokshitzer--Gribov--Lipatov--Altarelli--Parisi evolution equations for
QED. Analytical expressions for $\DD(w,\beta)$ are
known~\cite{Skrzypek:1992vk,Przybycien:1993qe,Arbuzov:1999cq} up to
the fifth order in $\alpha$ so that
terms ${\mathcal O}(\alpha^nL^n)$ can be found for $n=1,\ldots,5$
(in this paper we treat $n\leq 2$).

To find the various corrections outlined in the Introduction, we 
divide the SF into  three parts:  pure photonic, and non--singlet  (NS)
and singlet (S) $e^+e^-$ pair contributions,
\ba \label{dfun}
\DD(w,\beta) &\equiv & \DD_{\gamma}(w,\beta) 
+ \DD^{\mathrm{NS}}_{e^+e^-}(w,\beta)
+ \DD^{\mathrm{S}}_{e^+e^-}(w,\beta), 
\nonumber \\
\DD_{\gamma}(w,\beta) &=& \delta(1-w)
+ \sum_{n=1}\frac{\beta^n}{n!}\,P^{(n)}(w),
\nonumber \\
\DD^{\mathrm{NS}}_{e^+e^-}(w,\beta) &=& 
  \frac{\beta^2}{3}\, P^{(1)}(w) 
+ {\mathcal O}(\alpha^3L^3), 
\nonumber \\
\DD^{\mathrm{S}}_{e^+e^-}(w,\beta) &=& \frac{\beta^2}{2}\, R(w) 
+ {\mathcal O}(\alpha^3L^3),
\qquad
\beta \equiv \frac{\alpha}{2\pi}(L-1). 
\ea
The components  $\DD_{\gamma}$ and $\DD^{\mathrm{NS}}_{e^+e^-}$ 
correspond to those Feynman diagrams in which the registered positron
belongs to the same fermionic line as the initial one.
If the  registered positron arises from a pair production, 
its LL contribution to the energy spectrum is described by the
singlet function $\DD^{\mathrm{S}}_{e^+e^-}$. 
Functions relevant for our work are
\begin{eqnarray} 
P^{(n)}(w) &=& \lim_{\Delta\to 0}\biggl[
P^{(n)}_{\Delta}\delta(1-w) + P^{(n)}_{\Theta}\Theta(1-w-\Delta) \biggr],
\qquad 
\Theta(w) = 
\left\{\begin{array}{l} 1 \quad {\mathrm{for}} \quad w\geq 0 \\
0 \quad {\mathrm{for}} \quad w < 0
\end{array}\right. ,
\nonumber \\
P^{(1)}_{\Theta}(w) &=& \frac{1+w^2}{1-w}\, ,\qquad
P^{(1)}_{\Delta} = 2\ln\Delta + \frac{3}{2}\, , 
\nonumber \\ 
P^{(2)}_{\Theta}(w) &=& 2\biggl[ \frac{1+w^2}{1-w}\biggl(
2\ln(1-w) - \ln w + \frac{3}{2} \biggr) + \frac{1+w}{2}\ln w - 1 + w \biggr],
\qquad
P^{(2)}_{\Delta} = \biggl(2\ln\Delta + \frac{3}{2} \biggr)^2 
- \frac{2}{3}\pi^2,
\nonumber \\
R(w) &=& \frac{1-w}{3w}(4+7w+4w^2) + 2(1+w)\ln w.
\end{eqnarray}
Higher order expressions and further details
on the SF formalism can be found in 
\cite{Kuraev:1985hb,Kuraev:1997vn,Skrzypek:1992vk,%
Nicrosini:1987sm,Arbuzov:1999cq}.  

\begin{figure}[thbp]
\epsfig{file=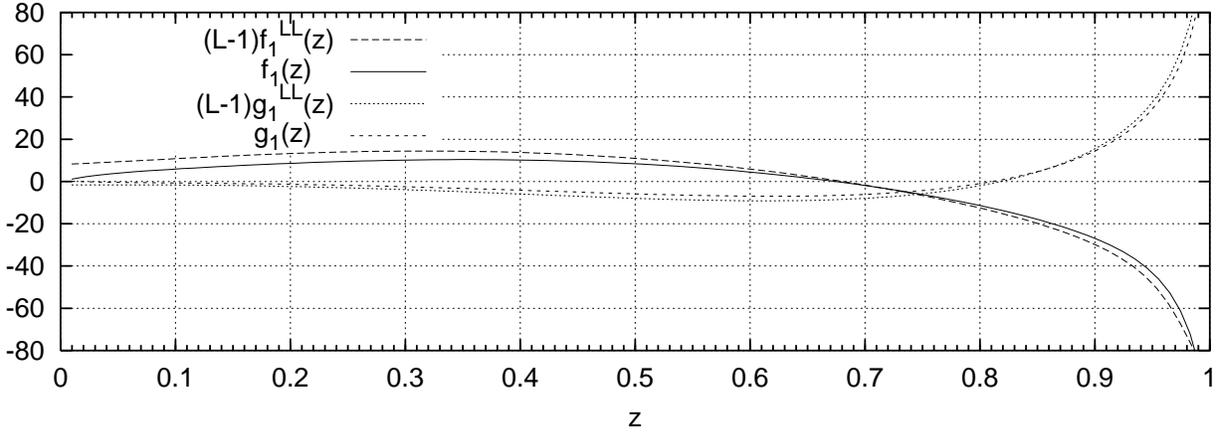,height=6cm} 
\caption{
Values of functions $f_1$ and $g_1$ versus $z$: 
exact results (eq.~(\ref{fgks})) and LL approximations 
(eqs.~(\ref{f1},\ref{g1})).}
\label{fg1fig}
\end{figure}

To find the LL corrections we use convolution, defined by
\begin{eqnarray} \label{conv}
A(\bullet)\otimes B(z) = \int\limits^1_0\dd w
\int\limits^1_0\dd w'\; \delta(z-ww')A(w)B(w') 
=\int\limits^1_z\frac{\dd w}{w}A(w)B\biggl(\frac{z}{w}\biggr).
\end{eqnarray}
For example, to reproduce the first order LL correction, we convolute
the Born--level spectrum with  $P^{(1)}$,
\ba \label{f1}
f_{1}^{\mathrm{LL}}(z) &=& P^{(1)}(\bullet)\otimes f_0(z)
= \frac{5}{6} + 2z - 4z^2 + \frac{8}{3}z^3
 +2z^2(3-2z)\ln{1-z\over z}, 
\\ \label{g1}
g_{1}^{\mathrm{LL}}(z) &=& 
 - \frac{1}{6} - 4z^2 + \frac{8}{3}z^3
 +2z^2(1-2z)\ln{1-z\over z}.
\ea
These formulas coincide with the LL parts of the full $\order{\alpha}$
results given in the Appendix.
A comparison of the LL and full first order functions is presented
in Fig.~\ref{fg1fig}. We see that the LL approximation
gives the bulk of the $\order{\alpha}$ correction, especially in 
the region of intermediate and large values of $z$, relevant for TWIST.

For the second order photonic LL corrections, we convolute with
$P^{(2)}$, 
\ba
f_{2}^{\mathrm{LL}(\gamma)}(z) &=& P^{(2)}(\bullet)
\otimes f_0(z)
\nonumber \\
&=& 4z^2(3-2z)\Phi(z) 
 + \biggl( \frac{10}{3} + 8z - 16z^2 + \frac{32}{3}z^3 \biggr)\ln(1-z)
\nonumber \\ 
&& + \biggl( - \frac{5}{6} - 2z + 8z^2 - \frac{32}{3}z^3\biggr)\ln z
+ \frac{11}{36} + \frac{17}{6}z + \frac{8}{3}z^2 - \frac{32}{9}z^3,
\label{f2gamma}
\\ 
g_{2}^{\mathrm{LL}(\gamma)}(z) 
&=& 4z^2(1-2z)\Phi(z)
+ \biggl( - \frac{2}{3} - 16z^2 + \frac{32}{3}z^3 \biggr)\ln(1-z)
\nonumber \\ 
&&+ \biggl( \frac{1}{6} + 8z^2 - \frac{32}{3}z^3\biggr)\ln z
- \frac{7}{36} - \frac{7}{6}z + \frac{8}{3}z^2 - \frac{32}{9}z^3,
\label{g2gamma}
\\ \nonumber
\Phi(z) &\equiv & \Li2{-{1-z\over z}} + \ln^2{1-z\over z}-{\pi^2\over 6},
\qquad
\qquad
\Li2{z} \equiv  -\int\limits_{0}^{z}\frac{\dd y}{y}\ln(1-y).
\ea
In the same manner we can get the third order photonic contributions.

Integrals of the LL photonic contributions vanish,
\ba \label{kln}
\int\limits_0^1 \dd z\; f_{1}^{\mathrm{LL}}(z) =
\int\limits_0^1 \dd z\; f_{2}^{\mathrm{LL}(\gamma)}(z) = 
\int\limits_0^1 \dd z\; g_{1}^{\mathrm{LL}}(z) =
\int\limits_0^1 \dd z\; g_{2}^{\mathrm{LL}(\gamma)}(z) = 0,
\ea
in accord with the theorem about the cancellation of mass
singularities \cite{Kinoshita:1962ur,Lee:1964is}.

\begin{figure}[thbp]
\epsfig{file=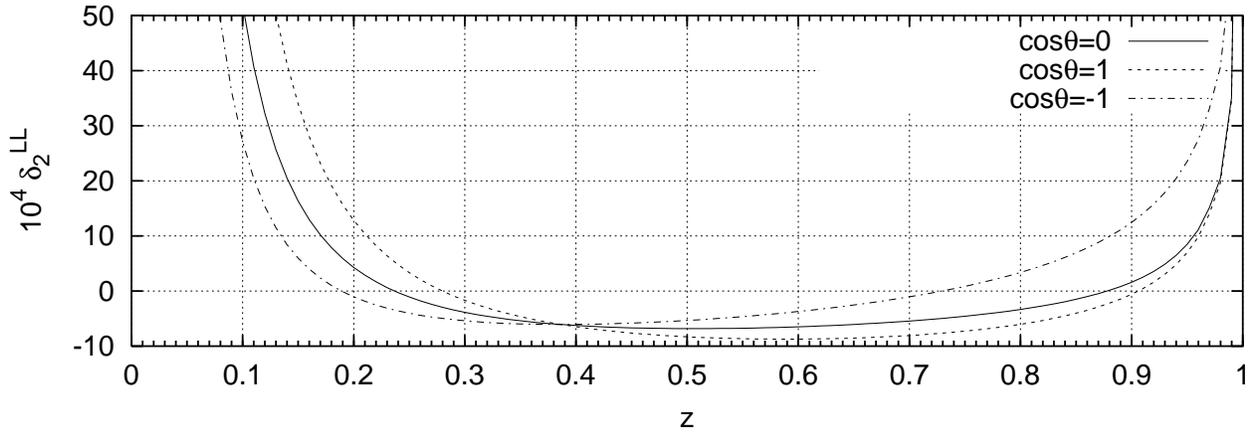,height=6cm} 
\caption{
Relative size of the second order LL photonic corrections for
as a function of $z$, plotted for three values of $\cos\theta$.
}
\label{d2gfig}
\end{figure}

A numerical illustration of our results for the relative size of the
second order LL photonic corrections is given in 
Fig.~\ref{d2gfig}, where we plot the relative correction defined as
\ba
\delta_{2}^{\mathrm{LL}(\gamma)} = \left(\frac{\alpha}{2\pi}\right)^2
\frac{(L-1)^2}{2} \left(\frac{f_{2}^{\mathrm{LL}(\gamma)}(z) 
- \cos\theta g_{2}^{\mathrm{LL}(\gamma)}(z)}
{f_{0}(z) - \cos\theta g_{0}(z)}\right).
\ea

In $\order{\alpha^2}$ the positron distribution has a contribution
from real and virtual $e^+e^-$ pairs.  Virtual effects of heavier
fermions are negligible~\cite{Davydychev:2001ee}. Pair correcttions are
found by convoluting $f_0$ and $g_0$ with $P^{(1)}$ (non--singlet)
and $R$ (singlet),
\ba 
f^{\mathrm{LL}(e^+e^-)}_{2\,\NS}(z) &=&     
f^{\mathrm{LL}}_{1}(z),
\qquad
g^{\mathrm{LL}(e^+e^-)}_{2\,\NS}(z) =
g^{\mathrm{LL}}_{1}(z)
\quad \mbox{(see eqs.~(\ref{f1},\ref{g1}))},
\label{ns2ee}
\\ \nonumber 
f^{\mathrm{LL}(e^+e^-)}_{2\,\mathrm{S}}(z)
&=& \frac{17}{9} + \frac{2}{3z}
 + 3z - \frac{14}{3}z^2 - \frac{8}{9}z^3
 + \biggl( \frac{5}{3} + 4z + 4z^2 \biggr)\ln z,
\\ 
g^{\mathrm{LL}(e^+e^-)}_{2\,\mathrm{S}}(z)
&=& -\frac{1}{9} - \frac{2}{9z}
 + z + \frac{2}{9}z^2 - \frac{8}{9}z^3
 + \biggl( -\frac{1}{3} + \frac{4}{3}z^2 \biggr)\ln z.
\label{g2ee}
\ea

In Table~I, numerical results are presented for the quantity
$\delta_{2}^{\mathrm{LL}(e^+e^-)}$, which gives the relative size
of the second order LL pair correction with respect to the Born
distribution (for $P_{\mu}=1$),
\ba
\delta_{2}^{\mathrm{LL}(e^+e^-)} &=&
\int\limits_{w_{\mathrm{min}}}^{1}\frac{\dd w}{w}\,
\frac{f_0(z/w)-\cos\theta g_0(z/w)}{f_0(z)-\cos\theta g_0(z)}
\biggl[ \DD^{\mathrm{NS}}_{e^+e^-}(w,\beta)
+ \DD^{\mathrm{S}}_{e^+e^-}(w,\beta)
\biggr], \qquad
w_{\mathrm{min}} = {\mathrm{max}}\left\{z,\frac{z}{z+y}\right\},
\ea
where $y$ is the cut on the maximal energy fraction of the real pair,
$E_{\mathrm{pair}} \leq y m_{\mu}/2$.
Both the singlet and the non--singlet pair contributions are
taken into account. It means that we simulated the situation, 
where an observation of two positrons is treated 
as a pair of simultaneous $\mu^+$ decays.
\begin{table}[htb]
\caption{Leading logarithmic pair correction 
$10^4\delta_{2}^{\mathrm{LL}(e^+e^-)}$ for
$\cos\theta=-1$.}
\begin{ruledtabular}
\begin{tabular}{rrrrrr} 
\multicolumn{1}{c}{$z\backslash y$} & 
\multicolumn{1}{c}{0.1}  &
\multicolumn{1}{c}{0.2}  &
\multicolumn{1}{c}{0.3}  &
\multicolumn{1}{c}{0.4}  &
\multicolumn{1}{c}{0.5}  \\ 
\hline
0.05 &   5.60 &  20.62 &  51.01 &  99.44 & 164.88 \\
 0.1 &   1.90 &   5.15 &  10.03 &  16.83 &  25.32 \\
 0.2 &   0.46 &   1.66 &   2.83 &   4.07 &   5.35 \\
 0.3 &$-$0.10 &   0.73 &   1.36 &   1.91 &   2.36 \\
 0.5 &$-$0.72 &$-$0.14 &   0.17 &   0.35 &   0.40 \\
 0.7 &$-$1.18 &$-$0.80 &$-$0.71 &$-$0.71 &$-$0.71 \\
 0.9 &$-$2.01 &$-$2.01 &$-$2.01 &$-$2.01 &$-$2.01 \\
\end{tabular}
\end{ruledtabular}
\label{pairtb}
\end{table}

In general, effects due to $e^+e^-$ pairs depend on
experimental conditions and cuts for events
with several charged particles in the final state.

If one is interested in the total LL effect due to real and virtual
$e^+e^-$ pairs to the muon decay width, one has to drop the singlet
pair contributions $f^{\mathrm{LL}(e^+e^-)}_{2\,\mathrm{S}}(z)$ and
$g^{\mathrm{LL}(e^+e^-)}_{2\,\mathrm{S}}(z)$, to avoid double
counting of real pairs.  Cancellation of the leading logarithms in
the total decay width is guaranteed by (\ref{kln}).

\section{Exponentiation}

Toward the energy spectrum end point $(z\to 1)$, the first order
correction $f_1^{\mathrm{LL}}$ and $g_1^{\mathrm{LL}}$ diverge.  This
phenomenon, discussed in
Refs.~\cite{Kinoshita:1959ru,Marciano:1975kw}, is a signal to look
beyond the first order approximation.  One can use the
Yennie--Frautschi--Suura theorem~\cite{Yennie:1961ad} to re--sum the
divergent terms and convert them into an exponential function.
Exponentiated representations of the SF can be employed to re--sum parts
of the leading logs to all orders in $\alpha L$
\cite{Kuraev:1985hb,Cacciari:1992pz}.  The exponentiation for the muon
decay has been criticized~\cite{Roos:1971mj}, because the large
logarithmic terms contain a mass singularity: not all large logs
disappear in the terms supplied by the exponent after an integration
over the energy, like in eq.~(\ref{kln}).

As has been discussed in \cite{Sachs72}, the validity of
exponentiation is limited to the region near the end of the spectrum.
One can see from the $\order{\alpha}$ results that the exponentiation
can be relevant only in a very small range where $z$ differs from 1 by
about $10^{-10}$ (the correction is about $-50\%$ at $1-z=10^{-10}$),
which is much less than the experimental resolution.  For this reason
we leave our results in the un--exponentiated,  fixed order form.

The end region of the spectrum is usually excluded from the fits of
Michel parameters. This is done to reduce the uncertainty due to
the finite energy resolution, which is most important in this region.
Indeed, the shape of an experimentally observed
spectrum is a convolution of the ``true'' spectrum with a resolution
function.  In the ``bulk'' part the effect of the finite resolution is
very small. However near a sharp edge (with the width much less than
the width of the resolution function) the shape of the convoluted
spectrum is defined mainly by the resolution function and not by the
spectrum.
Exclusion of the end point region helps also to reduce the theoretical
uncertainty because this is where the unknown higher--order
corrections are expected to be the largest.

\section{Conclusions}

To estimate the effect of the second order correction on values of
Michel parameters measured in an experiment, we generated
a 2D distribution in $z$
and $\cos\theta$ according to the RC-corrected spectrum
and fitted with a spectrum without the corrections.

$10^9$ toy Monte Carlo ``events'' $\{z,\cos\theta\}$ were produced by sampling the
2-dimensional spectrum~(\ref{spectrum}) taking into account
the complete first order corrections and the second order LL photonic ones.
This level of statistics is expected
to be accumulated in the TWIST experiment~\cite{Quraan:2000vq}.
The acceptance--rejection method and the Mersenne Twistor random number
generator~\cite{Mersenne-Twistor} was used.  
Events passing the ``acceptance cuts'' $0.34\le|\cos\theta|\le0.98$,
$0.4\le z$ and the cut $z\le z_{\mathrm{max}}$ were filled into a 2D
histogram.  $z_{\mathrm{max}}$ varied between 0.96--0.995.  The cuts
roughly represented acceptance of the TWIST
detector~\cite{Quraan:2000vq}. The region of $z$ close to 1 was
excluded to avoid the issue of the experimental resolution,
discussed above.  Finally, a maximum loglikelihood fit of the
spectrum {\em without} the second order RC to the
histogram was done.  $\rho$, $\eta$, $\xi$, $\delta$, and the global
normalization were the 5 free parameters of the fit. 
We put $P_\mu=1$ both in the generation and in the fits.
Binning of the histogram was chosen sufficiently fine, so that 
repeating the procedure with several times smaller bins 
gave the same results.
Self--consistency of the method was checked by fitting the histogram
with the full spectrum. Original values of Michel parameters were 
reproduced within the errors. 

We have observed statistically significant deviations of the fitted
Michel parameters from their original values when doing the fits
without the $f_2^{\mathrm{LL}(\gamma)}$ and $g_2^{\mathrm{LL}(\gamma)}$ 
contributions.  
That emphasizes the importance of using a precise enough 
theoretical spectrum shape for extracting values of Michel parameters 
in an experiment.

\begin{figure}[thbp]
\begin{tabular}{cc}
\epsfig{file=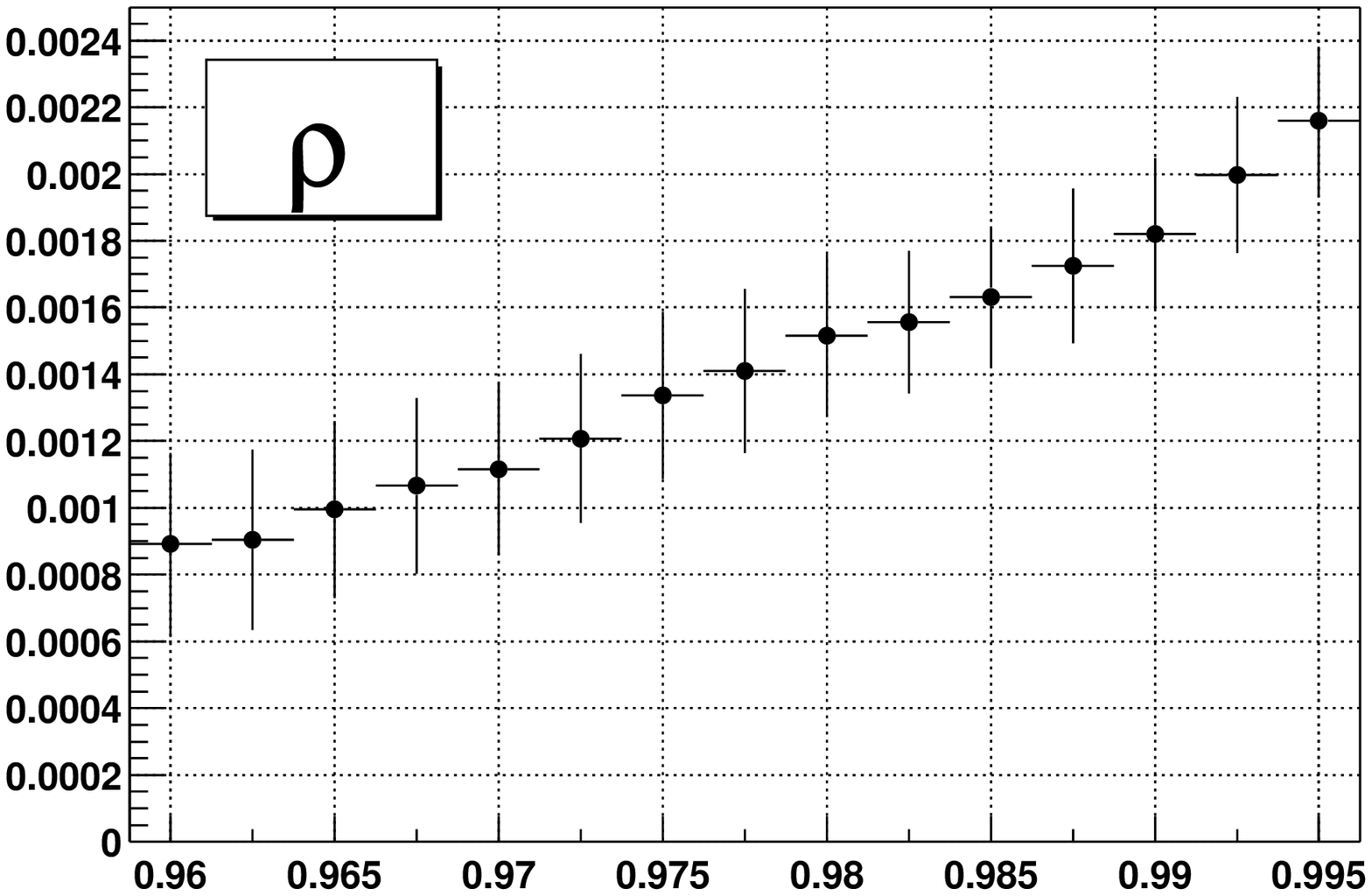,width=\splotwidth} 
& \epsfig{file=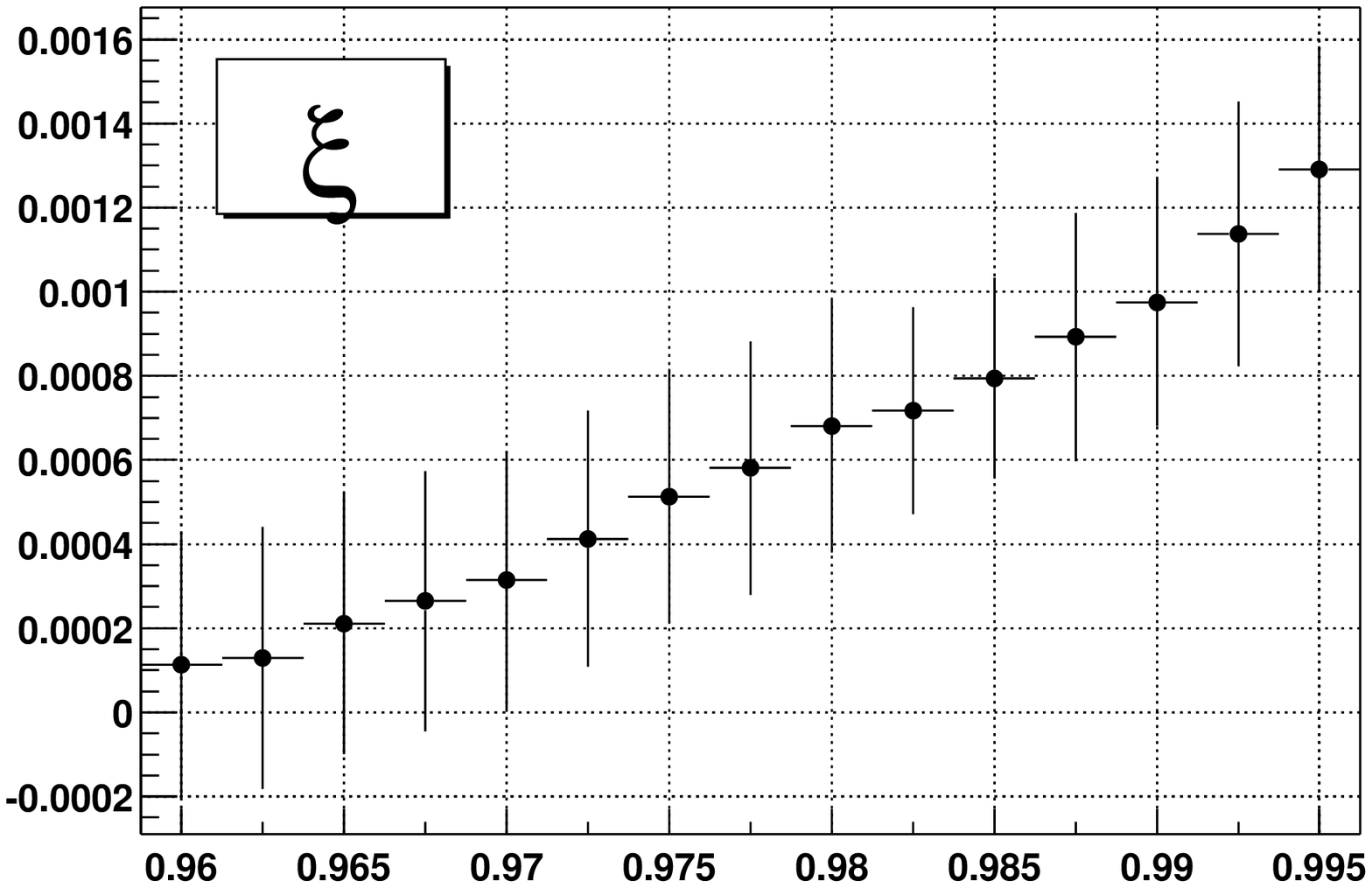,width=\splotwidth} 
\\
\epsfig{file=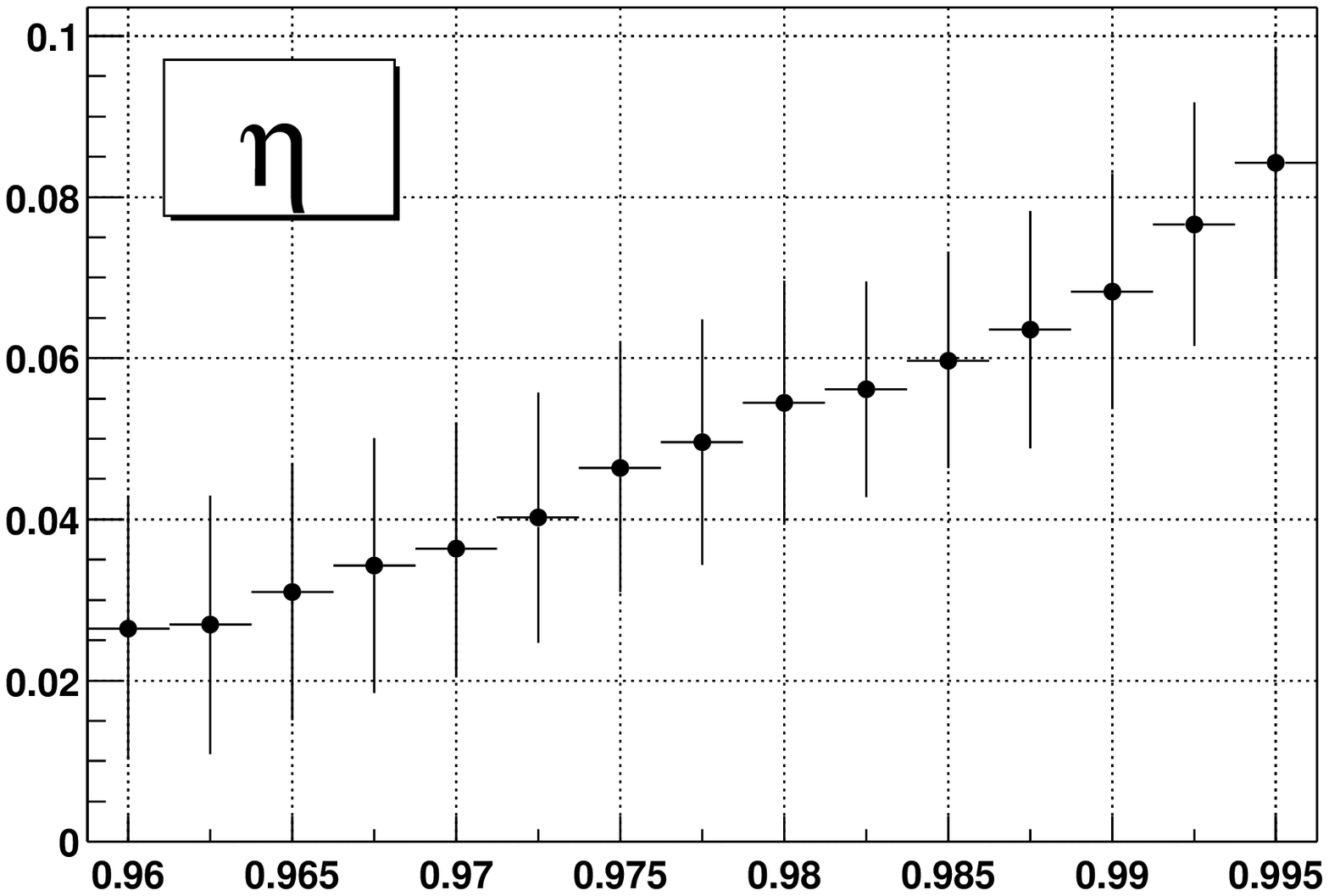,width=\splotwidth}
& \epsfig{file=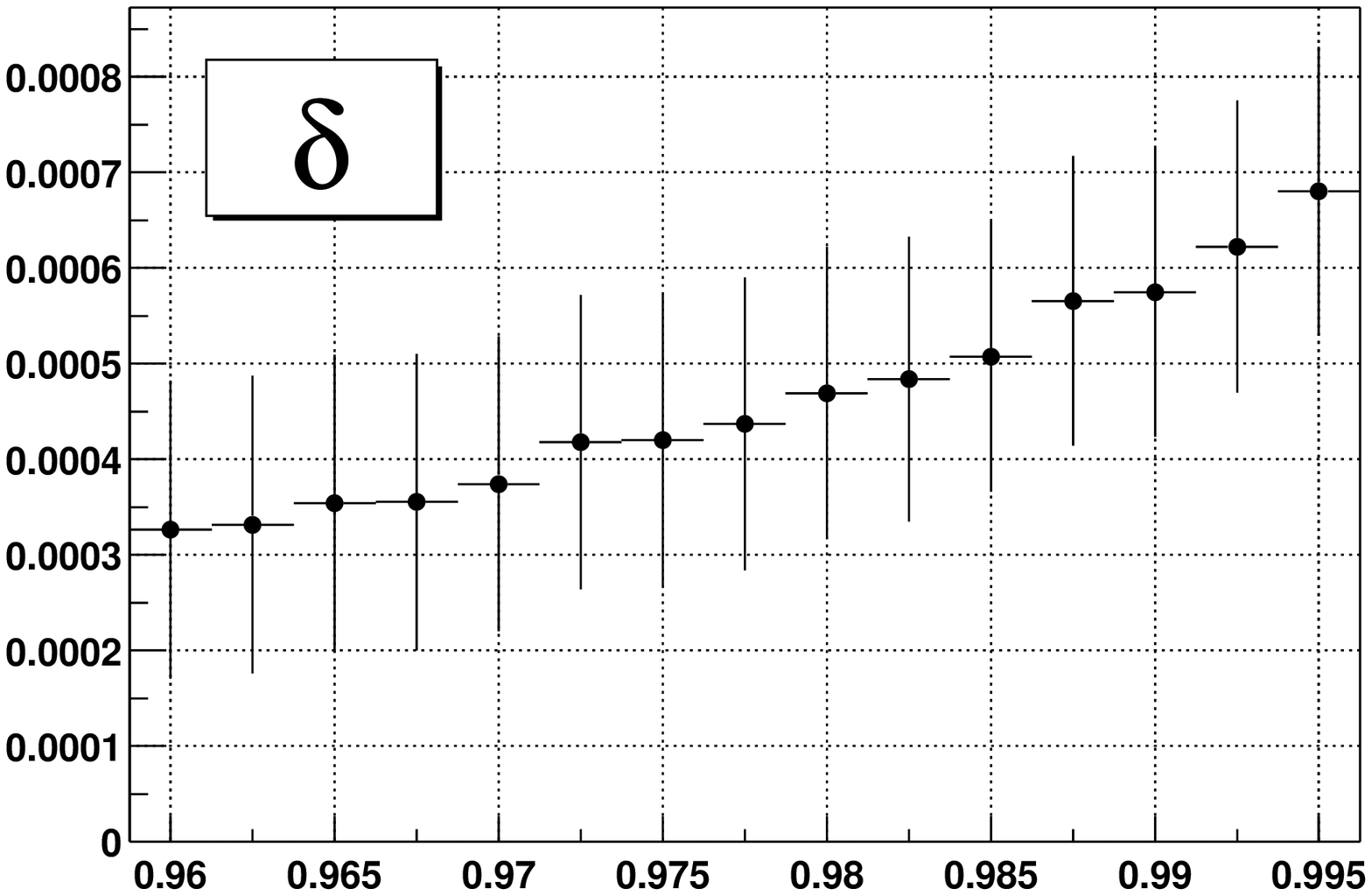,width=\splotwidth}
\end{tabular}
\caption{
Shifts of Michel parameters due to the second order
radiative correction for different upper energy cuts.
Horizontal axis is $z_{\mathrm{max}}$, vertical axis is the 
difference between a reconstructed parameter and its Standard Model
value.  The points are correlated since they are obtained from 
the same data set.
}
\label{rcfig}
\end{figure}

Shifts of Michel parameters due to radiative corrections depend on
the fit region, and perhaps on other factors not considered here. For
example, one may want to take into account the effect of the finite
experimental resolution in the bulk part of the spectrum.
Fig.~\ref{rcfig} demonstrates dependence of the shifts on the upper
energy limit of the fit region.

For a realistic value of the cut, $z_{\mathrm{max}}=0.97$,
the shifts of Michel parameters due to 
the second order LL corrections are of the order 
\ba
\Delta \rho & \simeq & 11\cdot10^{-4}, \nonumber \\
\Delta \eta & \simeq & 350\cdot10^{-4},\nonumber \\
\Delta \xi & \simeq & 3\cdot10^{-4}, \nonumber \\
\Delta \delta & \simeq & 4\cdot10^{-4}.
\label{eq:shifts}
\ea

The relatively large shift of $\eta$ is due to the fact that it
enters the spectrum with a small coefficient 
$z_0\simeq 10^{-2}$. 
If we fix $\eta$ during the fits to its SM value 0,
the shifts due to $\order{\alpha^2}$ effects become 
$5\cdot10^{-4}$ for $\rho$,
$-3\cdot10^{-4}$ for $\xi$,
and $3\cdot10^{-4}$ for $\delta$.
Comparing the results for free and fixed $\eta$ we observe a strong
correlation of $\eta$ with $\rho$ and $\xi$.

Fig.~\ref{rcfig} indicates a rather strong dependence of all Michel
parameters on $z_{\mathrm{max}}$.  It follows from the peaked
behavior of the radiative corrections when $z$ is close to $1$.  The
fitting procedure tries to compensate the effects of radiative
corrections and cuts by adjusting Michel parameters and the global
normalization. 

The remaining theoretical uncertainty in the muon decay
spectrum, eq.~(\ref{spectrum}), is due to unknown
contributions $\order{\alpha^2L}$
and $\order{\alpha^nL^n}$, $n\geq3$. Non--logarithmic terms of the
order $(\alpha/\pi)^2\simeq 5.4\cdot 10^{-6}$ are expected to be small
(compared with the $10^{-4}$ precision tag).  The sub--leading
contributions $\order{\alpha^2L}$ are the main source of the remaining
uncertainty. The magnitude of the corresponding photonic contribution
can be estimated using the known second order photonic LL correction,
\ba \label{nlla}
\delta_2^{\mathrm{NLL}(\gamma)} \sim 
\frac{3}{L}\delta_2^{\mathrm{LL}(\gamma)} \simeq
0.3\,\delta_2^{\mathrm{LL}(\gamma)}, 
\ea
where we multiplied by 3 to account for the unknown coefficient
of sub--leading terms.  
The leading contributions from higher orders $(n\geq 3)$ can be estimated
using again the second order LL result,
\ba \label{lla3}
\delta_3^{\mathrm{LL}(\gamma)} \sim 3\frac{\alpha}{2\pi}L
\delta_2^{\mathrm{LL}(\gamma)} \simeq 0.03\,\delta_2^{\mathrm{LL}(\gamma)}.
\ea 
$e^+e^-$ pair contributions are typically smaller than those of
the photons, at least in the LL.

These estimates of unknown radiative 
corrections can be converted into theoretical
uncertainties $\sigma^{\mathrm{th}}$ of Michel parameters, of the
order of a third of the 
shifts in eq.~(\ref{eq:shifts}) 
we found by including the LL photonic corrections.
With $z_{\mathrm{max}}=0.97$ we find
\ba \label{sigmath}
\sigma^{\mathrm{th}}(\rho ) &=& 3\cdot 10^{-4},
\nonumber \\
\sigma^{\mathrm{th}}(\eta ) &=& 100\cdot 10^{-4},
\nonumber \\
\sigma^{\mathrm{th}}(\xi P_{\mu}) &=&  1\cdot 10^{-4},
\nonumber \\
\sigma^{\mathrm{th}}(\delta ) &=&  1\cdot 10^{-4}.
\ea  
The conditions and the fitting procedure in a concrete experiment
can be different from the ones described above.
The actual size of the effect of radiative corrections 
on Michel parameters can be derived there in a similar way, 
starting from the analytical formulas for theoretical predictions 
and applying specific experimental conditions.

The planned accuracy of the TWIST
experiment~\cite{Rodning:2001js,Quraan:2000vq} is
\ba \label{sigmexp}
\sigma^{\mathrm{exp}}(\rho )       &=& 1\cdot 10^{-4},
\nonumber \\
\sigma^{\mathrm{exp}}(\eta )       &=& 30\cdot 10^{-4},
\nonumber \\
\sigma^{\mathrm{exp}}(\xi P_{\mu}) &=& 1.3\cdot 10^{-4},
\nonumber \\
\sigma^{\mathrm{exp}}(\delta )     &=& 1.4\cdot 10^{-4}.
\ea  

Clearly, effects of the second order LL radiative corrections,
eq.~(\ref{eq:shifts}),
have to be taken
into account at this level of accuracy.  In order to further reduce
the theoretical uncertainty, the next--to--leading second order
corrections should be evaluated as well.  Work on this is in progress.

\begin{acknowledgments}
This research was supported by the Natural Sciences and Engineering
Research Council of Canada, the Alberta Ingenuity Fund, and the
University of Alberta.  We thank Carl Gagliardi, Nathan Rodning, and
Vladimir Selivanov for valuable discussions concerning experimental
conditions and the fitting procedure.  
A.A. is grateful for the hospitality of the Brookhaven
National Laboratory and TRIUMF.

\end{acknowledgments}

\appendix*
\section{First order corrections}

$\order{\alpha}$ corrections to the muon decay spectrum, without  
terms suppressed by $m_e^2/m_\mu^2$, 
read \cite{Kinoshita:1959ru}
\ba \label{fgks}
f_{1}(z) &=& (L-1)f_1^{\mathrm{LL}}(z) + 2z^2(3-2z)R_1(z)
+ \frac{1-z}{6}\left[ ( 10 + 34z - 32z^2 )\ln z 
+ 5 - 27z + 34z^2 \right],
\\ \nonumber 
g_{1}(z) &=& (L-1)g_1^{\mathrm{LL}}(z) + 2z^2(1-2z)R_1(z)
-{1+27z^2-16z^3 \over 3}\ln z
 -\frac{1-z}{6}
\left( 7 - 13z - 30z^2 \right)
- \frac{4(1-z)^3}{3z}\ln(1-z),
\\ \nonumber 
R_1(z) &\equiv & - 2\Li2{1-z} + \ln z\ln(1-z) - 2\ln^2z 
- \frac{\ln(1-z)}{z} - \frac{5}{4}\, .
\ea
$f_1^{\mathrm{LL}}(z)$ and $g_1^{\mathrm{LL}}(z)$ are defined in
eqs.~(\ref{f1},\ref{g1}). 



\end{document}